# Gate-Tunable Memristive Phenomena Mediated by Grain Boundaries in Single-Layer MoS$_2$

Vinod K. Sangwan[1], Deep Jariwala[1], In Soo Kim[1], Kan-Sheng Chen[1], Tobin J. Marks[1,2], Lincoln J. Lauhon[1]*, and Mark C. Hersam[1,2]*

[1]Department of Materials Science and Engineering, Northwestern University, Evanston, Illinois 60208, USA.

[2]Department of Chemistry, Northwestern University, Evanston, Illinois 60208, USA.

*e-mail: lauhon@northwestern.edu, m-hersam@northwestern.edu

Continued progress in high-speed computing depends on breakthroughs in both materials synthesis and device architectures.[1, 2, 3, 4] The performance of logic and memory can be significantly enhanced by introducing a memristor,[5, 6] a two-terminal device with internal resistance that depends on the history of external bias voltage.[5, 6, 7] State-of-the-art memristors, based on metal-insulator-metal (MIM) structures with insulating oxides such as TiO$_2$, are limited by a lack of control over filament formation and external control of switching voltage.[3, 4, 6, 8, 9] Here, we report a new class of memristors based on grain boundaries in single-layer MoS$_2$ devices.[10, 11, 12] Specifically, the resistance of grain boundaries emerging from contacts can be easily and repeatedly modulated, with switching ratios up to ~10$^3$ and dynamic negative differential resistance. Furthermore, the atomically thin nature of MoS$_2$ enables tuning of the SET voltage by a third gate terminal in a field-effect geometry, providing new functionality not observed in other known memristive devices.

In this work, memristors were fabricated from monolayer MoS$_2$ films grown on oxidized Si substrates (300 nm SiO$_2$) by chemical vapor deposition (CVD) sulfurization of MoO$_3$ films (see Methods and Supplementary Section S1). Two Au electrodes on the MoS$_2$ define the memristor channel and an additional gate electrode (Si) is used to control the SET voltage (Fig. 1a). The best performing devices have grain boundaries (GBs) that are connected to only one of the two electrodes (Fig. 1a, Supplementary Fig. 2). We refer to such devices as intersecting-GB memristors. As-fabricated devices were preconditioned to a switchable state by an electroforming process that causes irreversible changes in current-voltage (*I-V*) characteristics (Supplementary Sections S2 and S3). Electroformed intersecting-GB memristors show a high resistance state (HRS) at $V = 0$ V that changes to a low resistance state (LRS) at high bias, following an abrupt increase in current (Fig. 1b, SET process, sweep a). The device stays in the LRS as the voltage is decreased to zero (sweep b). The ratio of resistance in the bistable states at zero bias $R_{HRS}/R_{LRS}$ is ~10$^3$ (Fig. 1c, inset). In the negative bias sweep, the device begins in the LRS (sweep c) and then changes to the HRS (RESET process, sweep d). The pinched hysteresis loop in *I-V* with $I = 0$ at $V = 0$ indicates the presence of a memristive element within the device.[5, 6, 7] Furthermore, the device shows a sudden increase in current (conductance $G \equiv \partial I/\partial V = 175$ μS) within a single bias step of 0.1 V (Fig. 1b). This current spike is followed by a dynamic negative differential resistance (NDR) that is commonly observed in memristive systems.[4, 13] This dynamic NDR depends on sweep rate and is therefore different from the static NDR observed in resonant devices.[4, 6] Sweeps a and d are described by $I \sim V^m$ ($V < 8.5$ V, Fig. 1c), where *m* increases monotonically with bias. This type of dependence is characteristic of space-charge limited current (SCLC), which has been observed in complex-oxide memristors[14] as well as in monolayer MoS$_2$ transistors.[15] The present MoS$_2$ memristors show 100× smaller SET fields ($\sim 10^4$ V/cm for $V_{SET}$ = 8.3 V in Fig. 1b) than those in conventional memristors, thereby promise switches and memories of low dynamic power (note that the standard operating point of $V = I = 0$ makes the static power of memristors inherently low).

Devices with different grain boundary orientations were analyzed to identify the switching mechanism. The bridge-GB memristor contains a grain boundary parallel to the channel and bridging the two electrodes (Figs. 1d,e). After electroforming (Supplementary Section S2), the bridge-GB memristor $I$-$V$ shows an extremely rapid current increase ($G > 1$ mS, Figs. 1d,e) followed by NDR that is larger than that observed in intersecting-GB memristors (Fig. 1b). In contrast to intersecting-GB memristors, bridge-GB memristors do not exhibit bistable states of different resistances at $V = 0$ V. Furthermore, the gate voltage ($V_g$) modulated current ratio decreases dramatically from $\sim 5\times 10^4$ in the pristine state to $\sim 2$ in the electroformed state (Supplementary Fig. 6c), which suggests the formation of a highly conducting filament between the electrodes. In contrast, the bisecting-GB memristor features a grain boundary perpendicular to the channel that does not contact either of the electrodes and possesses bipolar resistive switching that shows a broad current peak followed by a slow current decay in the NDR regime with $R_{HRS}/R_{LRS} \sim 4$ at zero bias (Fig. 1f; Supplementary Section S4).

The dependencies of these memristive phenomena on the grain boundary geometry, together with the absence of memristive behavior in control devices without grain boundaries (Supplementary Section S5) and repeatable multiple sweeps (Supplementary Section S6), suggest the following mechanism for the conductance modulation in intersecting-GB memristors. The electroforming process increases the overall resistance of the device in the OFF state by removing mobile dopants (identified below) from the region between the grain boundary tip and the opposing electrode. During the SET process, dopants migrate from the grain boundary region to the depleted region, thus increasing the conductance (ON). The dopants are driven away from the drain electrode and toward the grain boundary region during the RESET process, turning the device OFF. The power law *I-V* characteristics (as opposed to exponential *I-V*), two-terminal versus four-terminal measurements (Supplementary Sections S2), and comparisons between Au and Ti contacts (Supplementary Section S3) rule out Schottky barrier formation near contacts *via* anion segregation, as seen in TiO$_2$ memristors[9]. Direct evidence of grain boundary migration was found in a device containing both a bisecting-GB and a grain boundary connected to one of the electrodes (Fig. 2). After multiple sweeps at high biases (± 40 V), the bisecting grain boundary shifted by up to 3 μm. Similar significant motion of extended defects in 2D materials has been observed in atomic scale transmission electron microscopy (TEM) images of migrating dislocations in CVD-grown single-layer WS$_2$.[16]

Electrostatic force microscopy (EFM) and spatially-resolved photoluminescence (PL) spectroscopy are used to further elucidate the switching mechanism in the present devices (Fig. 3). In contrast to buried MIM memristors, *all-surface* monolayer MoS$_2$ memristors provide a new platform for *in-situ* probing of the underlying mechanisms in nanoionic switches. The abrupt change of the cantilever phase in the EFM images of Figs. 3c-e (and Supplementary Section S7) across a bisecting grain boundary (see Methods) indicates that the electrostatic potential drops primarily at the grain boundary (i.e., the grain boundary is resistive). This conclusion is consistent with the overall higher resistance of the bisecting-GB memristor compared to the bridge-GB memristor. However, sulfur vacancies in MoS$_2$ have been shown to accumulate near grain boundaries,[11, 17] and both the midgap states of sulfur vacancies[11, 18] and the electrons donated by dangling bonds should render the MoS$_2$ regions near grain boundaries *more* conductive. We therefore expect that the resistance of devices with grain boundaries can be modulated to the extent that sulfur vacancy concentrations in adjacent regions are modulated by vacancy migration. It follows that an electric field applied parallel to an intersecting-GB will promote migration of vacancies between the grain boundary and the depletion region. Circumstantial evidence of the key role of sulfur vacancies is provided by the fact that these memristive phenomena are only observed in devices that are intentionally grown to produce sulfur vacancies according to methods we recently reported.[19]

PL and Raman spectroscopy maps of electroformed devices provide additional evidence for higher concentrations of sulfur vacancies near grain boundaries (Figs. 3f,g; Supplementary Section S8). First, PL emission is enhanced near grain boundaries in the bridge-GB and intersecting-GB memristors (Figs. 3f,g; Supplementary Section S8), as was also reported previously[11] and associated with sulfur vacancies. In addition, the PL blue-shift and Raman blue-shift are consistent with $O_2$ chemisorption at sulfur vacancies[20] (Supplementary Section S8). Thus, spatially resolved analyses confirm that an increased density of defects exists near grain boundaries following electroforming. Sulfur vacancies are the most likely candidates for mobile anionic species within $MoS_2$. Mobile Mo cations are unlikely to be relevant due to the absence of an electrolytic medium or an electrochemically active metal.[4]

The proposed migration mechanism of mobile anions can also account for the switching characteristics in bridge-GB and bisecting-GB memristors. In bridge-GB memristors, anions segregate near the grain boundary during electroforming, which increases the conductivity of the boundary similar to vacancy migration in $ZnO$[21], $TiO_2$,[6, 9] and $TaO_x$ memristors.[22] The switching in Fig. 1e proceeds through a two-step process: first, the region near the grain boundary becomes highly conducting at high bias due to metallic-like transport across interacting defects at high density;[23, 24] second, Joule heating leads to a thermal rupture of the filament inducing NDR similar to threshold switching.[4, 13, 25] Thus, volatile switching and the NDR regime result from the deceasing width of the defect-rich region *via* the lateral drift of anions.[4] The absence of bistable states at $V = 0$ V in bridge-GB memristors is akin to complementary resistive switching in $TaO_x$ memristors.[26] In contrast, the broad current peak, NDR features, and asymmetric I-V characteristics[27] of bisecting-GB memristors are reminiscent of soft switching in non-filamentary $GaO_x$ memristors[28] and Cr-doped $SrZrO_3$ memristors[27]. The hysteretic *I-V* (Fig. 1g) can be explained by the dynamic and non-linear relation between drift (driven by electric field) and diffusion (driven by the concentration gradient) of anions forming depletion regions on either side of the bisecting grain boundary.[6, 7]

Finally, we show that the present 3-terminal $MoS_2$ memristors are gate tunable, which has not been observed in previous memristive systems and thus presents new opportunities for memristive circuits and related applications. In intersecting-GB memristors, the SET voltage ($V_{SET}$) can be varied from 3.5 to 8.0 V by varying the gate bias from 0 to 40 V (Fig. 4a), which suggests that independently addressable local gates in $MoS_2$ memristors could continuously adjust $V_{SET}$ to afford fault-tolerant architectures.[29] Furthermore, controlled $V_{SET}$ allows additional flexibility in designing complementary and bipolar resistive switching.[8, 1, 30] Bisecting-GB memristors offer an additional gate-tunable functionality, namely that the resistance values in each of the bistable states at zero bias can be controlled by more than three orders of magnitude (Fig. 4b), while the switching ratio (~4-6) and shape of the SET curves remain relatively unchanged with gate voltage. This adjustable resistance could be used as continuous-weight synapses in neuromorphic circuits.[29, 31] Independent control of device resistances can also lead to better uniformity and impedance matching between memristive circuits. Gate-tunable memristors further present opportunities for hybrid CMOS-memristor field-programmable architectures.[29, 32]

METHODS:

**Devices fabrication and measurement:** Monolayer MoS$_2$ flakes were grown by chemical vapor deposition.[11, 17, 19] The extent of sulfurization was controlled by heating the sulfur vapor to 150 °C and by restricting the time of exposure to sulfur vapor to 3 min. A higher sulfur vapor temperature (~170 °C) and a longer duration of exposure (10 min) results in stoichiometric triangular flakes.[19] As calculated from the shift in the threshold voltage, MoS$_2$ flakes grown at reduced sulfur vapor pressure show a defect-induced doping level of approximately ~ $1.7 \times 10^{12}$ cm$^{-2}$. Quantitative estimation of the stoichiometry at different growth conditions is reported in ref[19]. AFM and Raman spectroscopy were used to identify monolayer MoS$_2$ and grain boundaries (Supplementary Section S1). Devices were fabricated using electron-beam lithography, following a previously reported procedure.[33] Electrodes (70 nm Au/ 2 nm Ti) were deposited by thermal evaporation. After the lift-off process, the devices were submerged in N-methyl-2-pyrrolidone at 80 °C for 30 min to remove processing residues. All electroforming processes and electrical measurements were performed under vacuum (pressure < $2 \times 10^{-5}$ Torr) using a LakeShore CRX 4K probe station and Keithley 2400 source-meters.

**Scanning probe microscopy:** AFM and EFM scans were performed in an Asylum Cypher ES system. All AFM images were taken in tapping mode using NCHR tips (Nanoworld Inc). The resonance frequency of these cantilevers is ~300 KHz, and the nominal diameter of the tip apex is ~10 nm. Tapping mode imaging is operated in the repulsive regime by maintaining the phase signal below 90 degrees throughout the entire scan. For EFM imaging, the cantilever amplitude was set to the same value as tapping mode imaging while maintaining a distance of ~50 nm from the surface to avoid damage to the tip from tall electrodes (70 nm). EFM tips (NanoWorld PointProbe EFM) were monolithic Si coated with PtIr. The typical tip radius and resonant frequency were 25 nm and 75 kHz, respectively. EFM scans were captured on wire-bonded devices placed inside an inert environment cell consisting of continuously flowing pure nitrogen. Device electrical biasing was achieved during EFM using Keithley source-meters.

**Photoluminescence spectroscopy:** Micro-photoluminescence measurements were conducted using a confocal Raman system (WITec Alpha 300R) equipped with a 532 nm excitation source. The laser was focused using a 100× objective (NA = 0.9), and the power was kept below ~50 μW to avoid laser-induced sample heating/damage. Since the spatial resolution of the system is ~350 nm, a 300 nm step size was used for PL mapping. A 600 g/mm grating dispersed the photons prior to collection by a Si-based CCD camera.

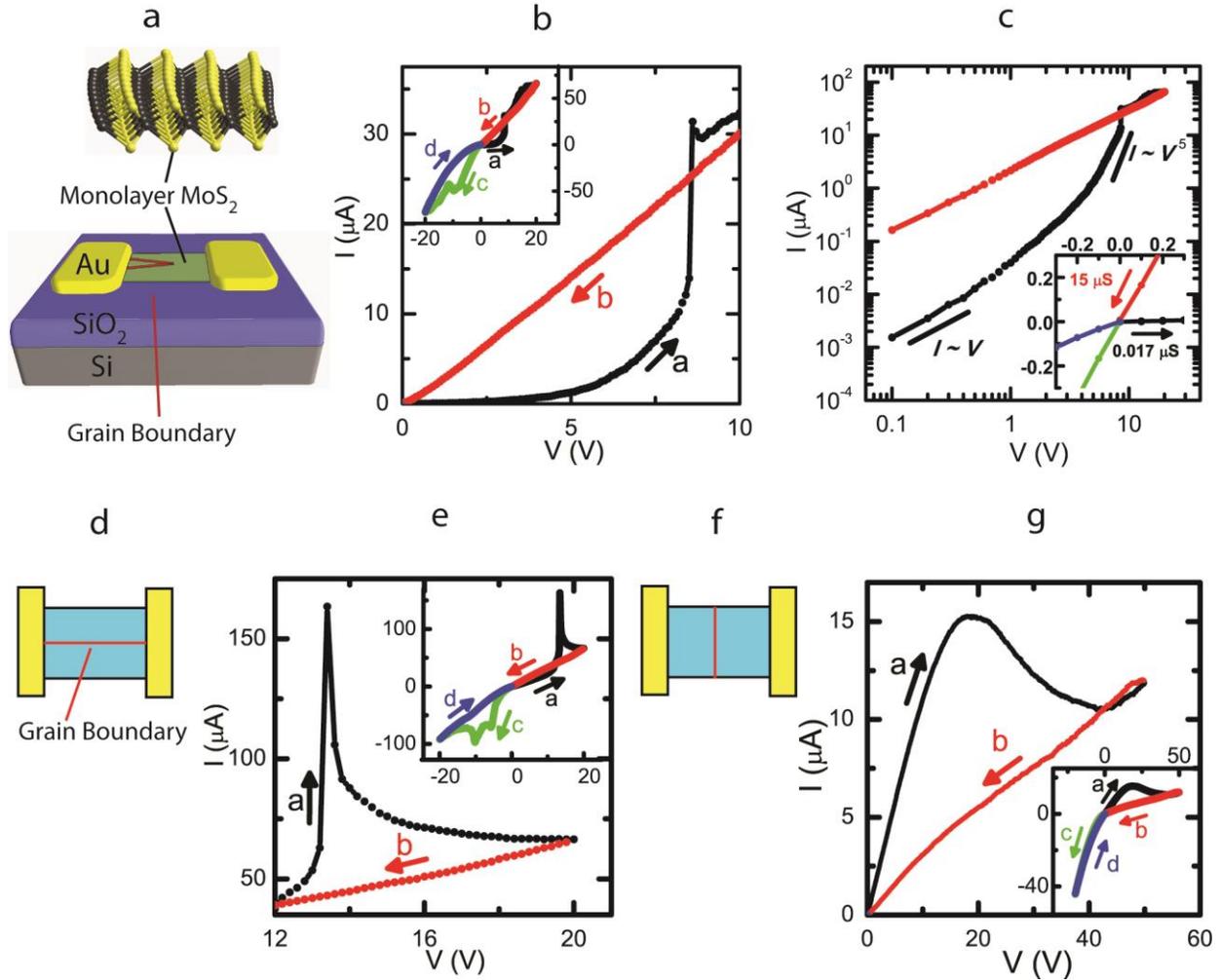

**Figure 1 | I-V characteristics of an MoS$_2$ memristor.** a) A schematic of an intersecting-GB MoS$_2$ memristor with two grain boundaries connected to one of the electrodes and intersecting at a vertex within the channel. b) Partial *I-V* characteristics of an electroformed intersecting-GB memristor (channel length $L$ = 7 μm; see Supplementary Section S1 for AFM images) obtained immediately after the electroforming process described in Supplementary Section S2. The SET process occurs at $V_{SET}$ = 8.3 V with abrupt two-fold increase in current. Inset shows full *I-V* characteristics of one switching cycle. Measurements were performed at a sweep rate of 1 V/sec and gate voltage $V_g$ = 40 V under vacuum (pressure < 2 × 10$^{-5}$ Torr). The voltage was swept in the order 0 V → 20 V → 0 V → -20 V → 0 V, as shown by the colored arrows with the four sweeps labeled as a, b, c, and d. c) Log-log plot of the sweeps a and b from Fig. 1b showing space-charge limited and ohmic transport in sweeps a and b, respectively. Sweep a also shows ohmic *I-V* behavior at low bias (*V* < 0.5 V). The inset shows a zoomed-in *I-V* curve from the

inset of Fig. 1b near zero bias, revealing conductance values $G$ ($\equiv \partial I/\partial V$) in OFF (HRS) and ON (LRS) states. d) A schematic and, e) partial *I-V* characteristic of an electroformed bridge-GB memristor ($L$ = 7.5 μm; see Fig. 3f and Supplementary Section S8 for AFM images) at $V_g$ = 40 V showing a transient current spike ($V_{SET}$ = 13.2 V) followed by an NDR regime. The inset shows full *I-V* characteristics of one switching cycle. The voltage sweeps are in the order a-b-c-d, as indicated by the colored arrows (sweep rate = 2 V/sec). f) A schematic and, g) partial *I-V* characteristic of an electroformed bisecting-GB memristor (see Supplementary Section S1 for an optical image and Supplementary Section S2 for electroforming process) at $V_g$ = 55 V showing a broad current peak followed by an NDR regime. The inset shows full *I-V* characteristics of one switching cycle. Bias sweeps were conducted in the order a-b-c-d, as indicated by the colored arrows (sweep rate = 2 V/sec).

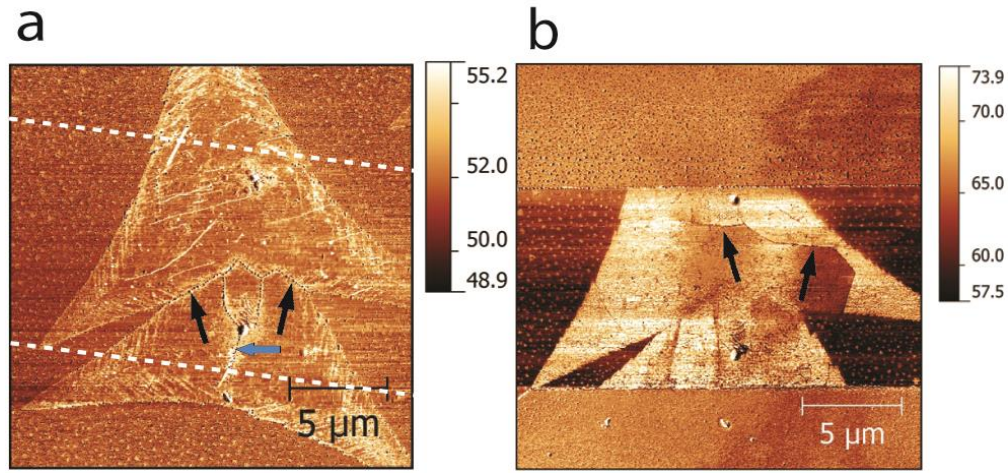

**Figure 2 | Grain boundary migration.** a) AFM phase image of an MoS$_2$ flake with multiple grain boundaries. The dashed white lines indicate the location of the electrode edges after device fabrication in (b). One grain boundary (highlighted by black arrows) is bisecting the channel while another grain boundary (blue arrow) touches the lower electrode edge. b) AFM phase image of the device after a series of 12 sweep cycles in the range of 40 V to –40 V (sweep rate = 1 V/s). Black arrows in (a) and (b) show that the grain boundaries have migrated by up to 3 µm. Color scale bars show the phase angles in degrees.

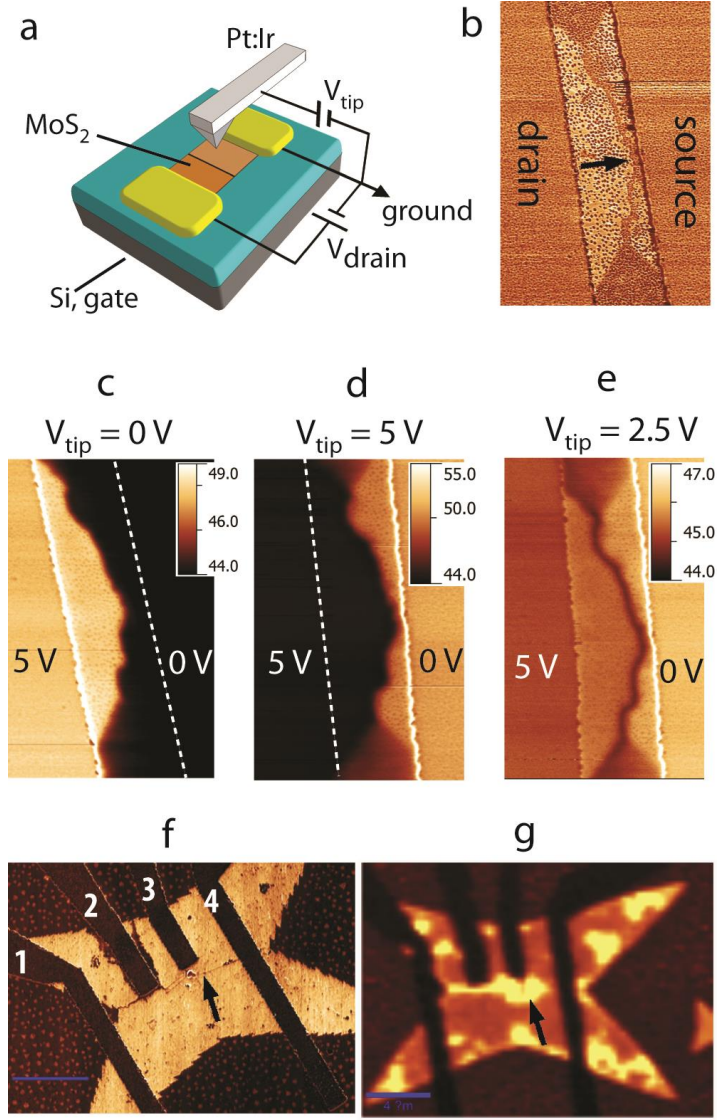

**Figure 3 | EFM and spatially-resolved PL images.** a) A schematic of the electrostatic force microscopy (EFM) measurement of the channel of a biased bisecting-GB memristor under an inert environment (see Methods). Bias voltages of the tip and drain electrode are varied while the source electrode and back gate Si are grounded. b) AFM phase image of an electroformed bisecting-GB device showing a grain boundary (highlighted by a black arrow) dividing the channel into two regions connected to 'drain' and 'source' electrodes, respectively. The channel length of the device is 2 μm. c), d), and e) Corresponding EFM phase images of the device from Fig. 3b at tip biases $V_{tip}$ = 5 V, 0 V, and 2.5 V, respectively. Color scale bars at the top right corners represent EFM phase in degrees. Device bias conditions are: $V_{drain}$ = 5 V and $V_{source}$ = $V_g$ = 0 V. The dotted lines highlight the metal-MoS$_2$ junctions with less contrast. f) AFM phase

image of an electroformed bridge-GB memristor with a grain boundary connecting both of the electrodes (highlight by a black arrow). Note that only the electrodes "1" and "4" of the van der Pauw geometry were used for electrical measurement while electrodes 2 and 3 were kept floating (Supplementary Section S2). g) Spatial mapping of the area under the photoluminescence (PL) excitonic peaks A and B (see Supplementary Section S8) of the $MoS_2$ device from Fig. 3f showing increased PL intensity in the grain boundary (blue arrow). The scale bars and Figs. 3f, g are 4 μm. See Supplementary Section S8 for correlated PL and Raman mapping and analyses.

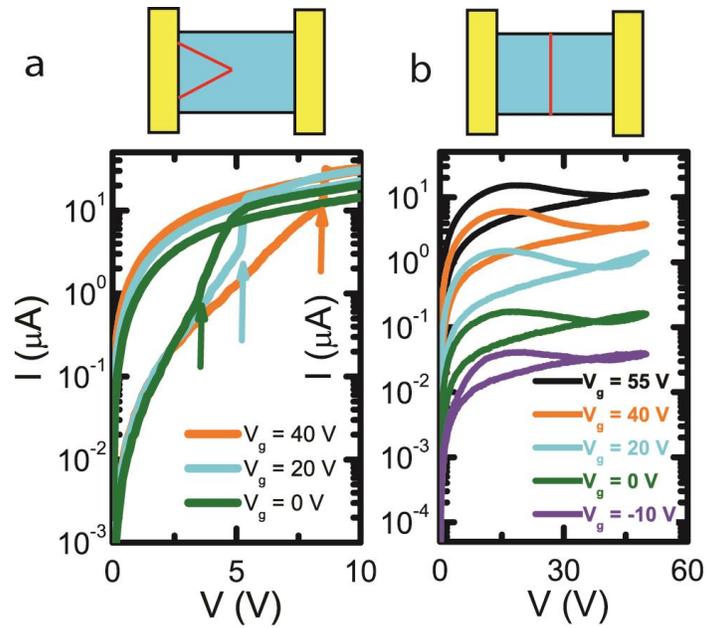

**Figure 4 | Gate-tunability of an intersecting-GB and a bisecting-GB memristor.** a) Log-linear plot showing *I-V* characteristics of an intersecting-GB memristor at different $V_g$. The full *V* sweep range is 20 to –20 V. Switching SET voltage (shown by colored arrows at $V_{SET}$ = 3.5, 5, and 8 V) is controlled by $V_g$. b) Log-linear plot of *I-V* characteristics ($V > 0$ V) of a bisecting-GB memristor at different $V_g$. The full *V* sweep range is 50 to –20 V. Negative bias sweep range is limited to –20 V to avoid dielectric breakdown between the gate and the drain electrode. Current in both the LRS and HRS is modulated by three orders of magnitude by the gate voltage.

ADDITIONAL INFORMATION

**Supplementary Information** accompanies this paper accompanies this paper at www.nature.com/naturenanotechnology.

**Competing financial interests**: The authors declare no competing financial interests.

**Reprints and permission information** is available online at http://npg.nature.com/reprintsandpermissions/.

**Correspondence** and requests for materials should be addressed to L.J.L. and M.C.H.

AUTHOR CONTRIBUTIONS

V.K.S., T.J.M., L.J.L., and M.C.H. designed the experiments. V.K.S. and D.J. fabricated and measured the devices. I.S.K. performed chemical vapor deposition, photoluminescence, and Raman microscopy. V.K.S. and K-S.C. conducted scanning probe microscopy (AFM/EFM) measurements. All authors wrote the manuscript and discussed the results at all stages.

ACKNOWLEDGMENTS

This research was supported by the Materials Research Science and Engineering Center (MRSEC) of Northwestern University (NSF DMR-1121262) and the Office of Naval Research (N00014-14-1-0669). This work made use of the EPIC facility (NUANCE Center, Northwestern University), which has received support from the MRSEC (NSF DMR-1121262), Nanoscale Science and Engineering Center (NSF EEC-0118025/003), State of Illinois, and Northwestern University.